\documentclass[a4paper]{mn2e}
\usepackage{epsfig}
\usepackage[
pdfauthor={M.E. Beer, A.R. King, M. Livio and J.E. Pringle},
pdftitle={How special is the Solar System?},
pdfsubject={We argue that, in terms of those planetary systems which
  have been observed, the case for the solar system being a typical
  planetary system has yet to be established},
pdfkeywords={accretion, accretion discs - planetary systems: protoplanetary discs}
]{hyperref}

\title[How special is the Solar System?]{How special is the Solar System?}

\author[M.\,E. Beer, A.\,R. King, M. Livio \& 
  J.\,E.~Pringle]{M.\,E. Beer$^1$\thanks{E-mail: martin.beer@astro.le.ac.uk}, A.\,R. King$^1$, M. Livio$^2$ and
  J.\,E. Pringle$^{2,3}$\\ 
$^1$Theoretical Astrophysics Group, University of Leicester, Leicester,
  LE1 7RH, UK\\ 
$^2$Space Telescope Science Institute, 3700 San Martin Drive,
  Baltimore, MD21218, USA\\ 
$^3$Institute of Astronomy, Madingley Road, Cambridge, CB3 0HA, UK}

\date{Accepted 2004 July 21st. Received 2004 June 30th; in original
  form 2004 March 29} 

\volume{000}

\pagerange{\pageref{firstpage}--\pageref{lastpage}} \pubyear{2004}

\begin{document}
\label{firstpage}
\maketitle

\begin{abstract}

Most mechanisms proposed for the formation of planets are modified
versions of the mechanism proposed for the solar system.  Here we
argue that, in terms of those planetary systems which have been
observed, the case for the solar system being a typical planetary
system has yet to be established. We consider the possibility that most
observed planetary systems have been formed in some quite different
way. If so, it may be that none of the observed planetary systems is
likely to harbour an earth-like planet.

\end{abstract}


\section{Introduction}
\begin{figure}
\begin{center}
\epsfig{file=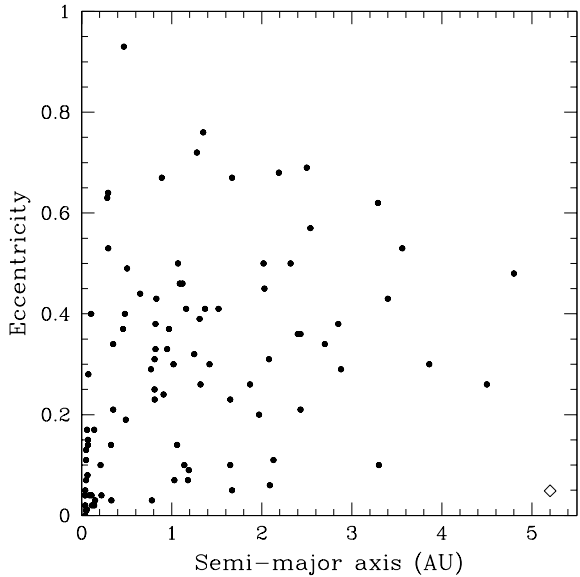,width=8cm}
\caption{A plot of eccentricity versus semi-major axis for the sample
discussed in this paper. The diamond at the bottom right represents
the solar system in the form of Jupiter.}
\label{eversusa}
\end{center}
\end{figure}

Until about ten years ago, theories of planetary formation
concentrated, of necessity, on the formation of planets in the solar
system (e.g. Lissauer 1993). The theories provide a reasonable
explanation, in terms extending back to the ideas of Kant (1755) and
others, of why the gas giants form at radii of more than around 5AU
from the sun, of why the basic building block of a planet is a rocky
core, built up of dust via planetesimals in the solar nebula, and of
why the orbits of the planets are essentially coplanar and
circular. In the last ten years, over a hundred planetary systems have
been discovered, and, at first glance, none of them resembles the
solar system (or what the solar system would appear to be when seen
from an appropriate distance). In comparing the various systems, it
seems sensible as a first step to just consider the properties of the
planet which has the largest velocity semi-amplitude in each observed
system. In Figure~\ref{eversusa} 
we plot the observed eccentricity, $e$, against the semi-major axis,
$a$, (measured in AU) for the planet with the largest velocity
semi-amplitude in each of the
observed systems, including Jupiter as the representative for the
solar system. (We discuss the sample used in our analysis in
Section~\ref{perianalysis} and we find no bias in the choice of this
as our sample).  It is apparent from Fig.~\ref{eversusa}
that in the observed systems the planets are much closer to the
central star (indeed too close to the central star to have formed by
the conventional theory -- Lin, Bodenheimer \& Richardson 1996;
Bodenheimer, Hubickyj \& Lissauer 2000), and have in general,
compared to the solar system, highly eccentric orbits (except for
those close enough to the central star that tidal circularization has
had time to act -- Rasio et al. 1994). Looked at as a distribution in
the two-dimensional $(a,e)$-plane it is immediately apparent that the
solar system does not lie within the part of the $(a,e)$-plane defined
by the currently observed distribution of exo-planetary systems.

Thus the natural question arises of whether the solar system is
special in some way compared to the majority of planetary systems to
be found in the Galaxy (c.f. the discussion in Ford, Havlickova \&
Rasio 2001). Of course, we know already that it is special in the
sense that we are in it, and it is therefore no surprise that all the
early models of the formation of planetary systems applied entirely to
the solar system.\footnote{Just as almost all models in cosmology apply
entirely to the Universe.}  But if the solar system is strongly
atypical as a planetary system, in the sense of lying at some extreme
in the distribution of properties of all planetary systems, then the
more relevant question arises. To what extent we should allow our
understanding of the formation of planetary systems, gleaned from
observing one atypical and extreme example, albeit at close-range, to
bias our understanding of the formation of planetary systems in
general?

In Section 2 we quantify the extent to which the solar system may, on
current evidence, be considered to be a typical member of the
ensemble of planetary systems. We make use of a technique derived in
the study of robust statistical procedures in order to determine
whether or not to discard some of the observations because they are
inconsistent with the rest of the observations and/or with the
probability distribution assumed to be the underlying distribution of
the data. We show that the solar system is, formally, an outlier from
the rest the distribution defined by the observed exo-planetary
systems. In Section~3 we discuss possible reasons for this result, and
address the question of whether it is currently justified to allow the
habitability of the solar system to influence our views on planet
formation in general. We summarise our conclusions in Section~4.

\section{Robust statistics}
In this Section we address the formal question of whether the solar
system, as represented by Jupiter, is an outlier from the observed
distribution of exo-planetary systems.  The properties of the known
extrasolar planets are distributed non-parametrically (i.e. there is
no known underlying distribution for the data). Consequently we
require a non-parametric statistical method in order to identify any
outliers which may exist in the data. Once any outliers have been
identified and removed, the remaining dataset may be described in a
statistical sense, as robust. In order to identify outliers in a
dataset, a standard procedure is to use a boxplot (see, for example,
Hoaglin, Mosteller \& Tukey 1983; Madansky 1988). However, in order
to quantify the degree to which a particular point is an outlier, the
simple boxplot does not suffice.  In order to do this, one standard
procedure is to transform the data so that it appears more closely to
resemble a normal distribution. One class of such transformations is
the power transformation as described below.  Since the solar system
(Jupiter) differs from the bulk of the observed systems both in terms
of large semi-major axis, $a$, and small eccentricity, $e$, we apply
our statistical considerations to the observed distribution of
periastron distance, $a_p = a(1-e)$.

\subsection{Box-Cox power transformation} \label{sectboxcox}
The Box-Cox power transformation (Box \& Cox 1964; see chapter 5 of
Madansky 1988 for a detailed discussion) is used to transform a set of
observed data points so that they are more normally distributed. The
Box-Cox transformation is of the form
\begin{equation}
y_{\lambda} (x) = \left \{ \begin{array}{ll}
\frac{\rule[-2pt]{0pt}{0pt}\hbox{$x^{\lambda} -
1$}}{\rule{0pt}{8pt}\hbox{$\lambda$}} & \rm{if~} \lambda \ne 0 \\
\log x & \rm{~~~} \lambda = 0
\end{array} \right .
\end{equation}
and $y_{\lambda} (x)$ will have a normal distribution exactly only if
$\lambda=0$ or $1/\lambda$ is an even integer. The aim of the
Box-Cox transformation is to find the value of $\lambda$ which gives
the best approximation to a normal distribution for the transformed
data. If $y_{\lambda} (x)$ is normally distributed the
maximum likelihood estimator of the mean of the transformed data is 
\begin{equation}
\bar{y}_{\lambda} = \sum^n_{i=1} \left . \frac{y_{\lambda i}}{n}
\right . ~,
\end{equation}
and the maximum likelihood estimator of the variance is
\begin{equation}
s^2_{\lambda} = \sum^n_{i=1} \left . \frac{ (y_{\lambda i} -
  \bar{y}_{\lambda}) ^2 }{n} \right . ~.
\end{equation}
The log likelihood function is 
\begin{equation}
l(\lambda) = -\frac{n}{2} \log (2\pi) - \frac{n}{2} - \frac{n}{2}\log
  s^2_{\lambda} + (\lambda-1)\sum^n_{i=1} \left . \log x_i \right .~.
\label{eqnloglike}
\end{equation} 
We may vary $\lambda$ to find the value at which this is a maximum and
then look for outliers in the transformed distribution.

\subsection{Periastron analysis}
\label{perianalysis}
We perform the Box-Cox analysis on the periastron distribution of the
known extrasolar planets. The data for this analysis was taken from
the California and Carnegie planet search catalogue
(http://exoplanets.org/planet\_table.shtml)\footnote{The parameters for
HD~190360 (also known as Gl777A) which had been claimed to be the
extrasolar planetary system most resembling Jupiter have recently been
updated (Naef et al. 2003) from those originally
published (Udry, Queloz \& Mayor 2003). The semi-major axis has
increased from 3.65\,AU to 4.8\,AU and the eccentricity from 0.0 to
0.48.}. In this analysis we only consider companions from the
extrasolar planet catalogue which have the largest velocity
semi-amplitude (i.e. were discovered first in most cases). We have
also performed our analysis using both the most massive extrasolar 
planets in each system and using all the known planets but the
significance of Jupiter as an outlier does not vary.

\begin{figure}
\begin{center}
\epsfig{file=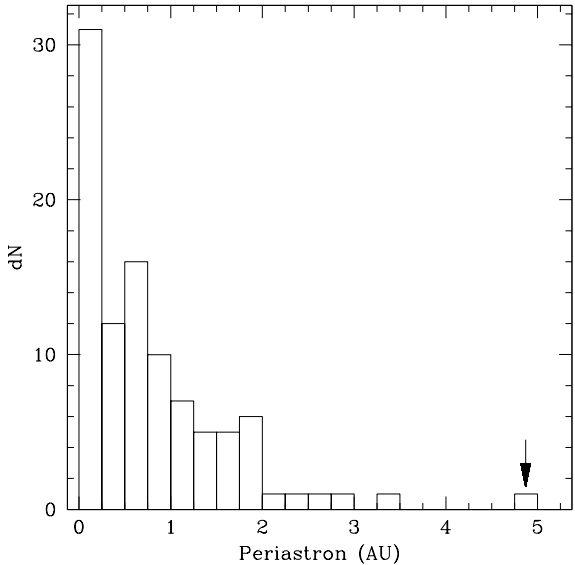,width=8cm}
\caption{A histogram showing the periastron distribution of the 
  extrasolar planets in our sample and Jupiter. The arrow indicates the bin
  containing Jupiter.} \label{figpretransform}
\end{center}
\end{figure}

Fig.~\ref{figpretransform} shows the
distribution of periastron distances of the known extrasolar planets
plus Jupiter. In Fig.~\ref{figpretransform} it is clear that Jupiter
(with a periastron distance of just under 5 AU) is an outlier but how
significant an outlier is unclear as the underlying distribution is
unknown. 

\begin{figure}
\begin{center}
\epsfig{file=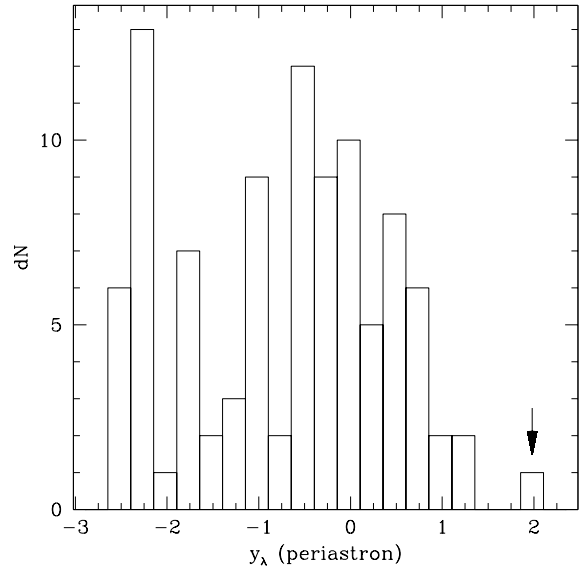,width=8cm}
\caption{A histogram showing the periastron distribution of the
extrasolar planets in our sample and Jupiter after a Box-Cox power
transformation 
has been applied as described in section~\ref{sectboxcox}. The arrow
indicates the bin containing Jupiter.}
\label{figposttransform}
\end{center}
\end{figure}

A Box-Cox transformation finds that $\lambda =0.185$ gives the maximum for
the log likelihood function
(equation~\ref{eqnloglike}). Fig.~\ref{figposttransform} shows the 
transformed data. We can see how well this resembles a normal
distribution by measuring the skewness and kurtosis of the transformed
data. The skewness is given by
\begin{equation}
S = \sum^n_{i=1} \left. \frac {1}{n} \left ( \frac{y_{\lambda i} -
  \bar{y}_{\lambda}} {\sigma} \right)^3 \right .~,
\end{equation}
and the kurtosis by
\begin{equation}
K = -3 + \sum^n_{i=1} \left. \frac {1}{n} \left ( \frac{y_{\lambda i} -
  \bar{y}_{\lambda}} {\sigma} \right)^4 \right .~,
\end{equation}
where $\sigma$ is the standard deviation of the distribution (see
Press et al. 1996, p.~606).  The
transformed data has a skewness of $-0.09$ and a kurtosis of $-1.07$. For
a normal distribution of $n$ data points we would expect a skewness of
$\sqrt{6/n}$ and a kurtosis of $\sqrt{24/n}$. Our dataset has 98
points so we would expect a skewness and kurtosis of up to 0.25 and
0.49 respectively. Clearly the transformed data is not skewed but it
does have a significant kurtosis. This is caused by the well-known
spike near the lower end of the distribution, consisting of `hot
Jupiters' at small orbital radii (Udry et al. 2003).

The mean and standard deviation of this transformed data are $-0.71$ and
1.11 respectively. Jupiter is a significant outlier at the $2.31\,\sigma$
level and is the only two sigma data point in the entire
distribution. Jupiter has a transformed periastron measurement of 1.86
while the next largest is that of HD~72659 which has a transformed
value of 1.35 i.e. it is half a sigma from its nearest neighbour. Note
that the fact that we have used the AU as our unit of measurement,
does not affect the outcome. For example, we have also applied the
Box-Cox transformation to the periastron distance measured in SI units
and to the logarithm of the periastron distance and find that the
result is the same in all cases with the transformed data resembling
that of Fig.~\ref{figposttransform} (although with different values
for $\lambda$). 

\section{Discussion}

We have noted that the solar system is an outlier in the parameter
space of known planetary systems. This is already evident from
Fig.~\ref{eversusa}, and we have used a non-parametric statistical
test to show that 
Jupiter lies at about 2.3\,$\sigma$ in the tail of the periastron
distribution. Given this, it seems sensible to enquire whether the
solar system is a typical planetary system in the sense that it
actually belongs to the set of exo-planetary systems observed so far,
or whether it might be intrinsically different, or special, in some
way. In particular, we address the question of the extent to which it
is necessary to assume that all planetary systems formed in the same
way as the solar system. The major issue here is whether theoretical
models for the formation of exo-planetary systems should just consider
minor modifications of the current theory for the formation of the
solar system, or whether some quite different ideas might need to be
investigated.

\subsection{Selection effects}
The most obvious point to make is that the way in which the solar
system was discovered differs from the rest. Nevertheless, it can be
argued that systems similar to the solar system are abundant, but have
not yet been discovered due to observational selection effects. This
point is made by Lineweaver \& Grether (2002, 2003) and by Tabachnik
\& Tremaine (2002). These authors fit power-law distributions to the
properties of extra-solar planets, and then extrapolate them to
encompass the properties of the solar system to make deductions about
the total number of planetary systems likely to be present, as well as
their properties. However, the act of fitting a power-law distribution
makes the implicit assumption that the properties of planetary systems
are essentially scale-free, and the act of extrapolating the
distribution to encompass the solar system (and beyond) makes implicit
assumptions about the general properties of planetary
systems. Lineweaver \& Grether (2002) emphasise that overcoming the
selection effects is mainly a matter of time. In other words, it may
be that all that is required to find systems like the solar system is
to observe candidate systems for timescales as long as Jupiter's
orbit.  This will take another 5 to 10 years. Thus application of a
test of the kind used above to the dataset some 5 to 10 years from now
will demonstrate the credibility of this potential solution.

\subsection{Dichotomy}

There is, however, already some evidence that the formation of
planetary systems is not a scale-free process. Greaves et al. (2004a)
find that there is a dichotomy between those systems which are
observed to contain giant planets (at radii typically a few AU or
less) and those systems which have a debris disc as evidenced by
submillimetre emission at radii typically in the range 20 -- 100
AU. They find that those systems with observed giant planets show no
evidence for a debris disc, whereas, those systems which display
evidence for a debris disc are not directly observed to harbour giant
planets. However, as Greaves et al. (2004a) point out, there is
evidence that these latter systems contain (currently unobserved)
planets at larger radii. This evidence is in two forms. First, the
absence of a mid-infrared excess in the debris disc systems points to
the presence of planets orbiting at larger radii, which are able to
clear a central cavity in the dust distribution, and second, the
suggestion of planets at larger radii is given credence by the
evidence of internal structure seen in those debris discs which can be
imaged, such as clumping, gaps and warping (Greaves et al. 1998;
Jayawardhana et al. 1998; Weinberger et al. 1999; Wyatt et al.
1999; Lagrange, Backman \& Artymowicz 2000; Augereau et al. 2001;
Holland et al. 2003; Wyatt 2003; Kalas, Liu \& Matthews
2004). Meyer et al. (2004) recently demonstrated the power of the Spitzer
Space Telescope at investigating the properties of debris discs. They
note that it is already clear that the interpretation of such discs is
likely to be complicated and to involve a number of factors including
perhaps a range in primordial disc properties and/or evolutionary
histories. 

\subsection{Theoretical models}

We therefore consider the implications of this observed break in
scale. It might for example be an indication that there are two quite
separate mechanisms of planet formation, or it might simply indicate
that we are looking at two extremes of a single formation process.

The majority of the current thought (and literature) on the formation
of planetary systems (actually giant planets) makes use of the
formation model derived for the solar system, modified in some
manner. This standard model for the solar system (Wetherill 1980;
Mizuno 1980; Stevenson 1982) assumes that planets form initially
through the agglomeration of dust into grains, pebbles, rocks and
thence planetesimals within a gaseous disc, that these planetesimals
coalesce to form planetary cores, and that finally (for the giant
planets) these cores use gravity to accrete gas from the ever-present
disc. Most of the current theoretical research effort on exo-planets
involves devising and justifying suitable modifications of this model,
taking the overall view that all planetary systems form a continuous
distribution in some suitable parameter space. Even for the solar
system, however, this scenario is not without problems (Lissauer
1993; Wuchterl, Guillot \& Lissauer 2000). Most notably, the
formation of planetesimals from small-scale dust, and the
timescales required to accrete the gas onto the proto-planetary cores,
encounter difficulties, 
given possible rapid migration of the cores (Ward \& Hahn 2000) and
the required core masses for runaway accretion, compared to those
observed (see also the discussion by Boss 2000). The main
modifications required to make the standard model fit the observed
systems, involving planetary migration (e.g. Trilling, Lunine \& Benz
2002; Armitage et al. 2002; Alibert, Mordasini \& Benz 2004) and the generation of
orbital eccentricities (e.g. Rasio \& Ford 1996), have difficulties
which are not yet fully resolved (see, for example, the discussions by
Marcy et al. 1999; Udry, Mayor \& Santos 2003; Tremaine \& Zakamska
2004).

The main alternative scenario involves the formation of giant planets
directly, through gravitational instability in the protostellar gaseous
disc (Boss 2001; Rice et al. 2003; Mayer et al. 2002; 2004). Although this
model has problems fitting the solar system (for example, the presence
of rocky cores in the giant solar system planets, not to mention the
existence of terrestrial planets, is more simply explained through the
prior formation of the cores from rocky planetesimals), there is no
fundamental reason why it, or some modification of it, should not
apply to the rest -- that is, to the majority of the systems
discovered so far. The major problem with the model, as currently
constructed, is that, in order to drive strong gravitational
instability, it requires a sudden change in disc properties on a
timescale comparable to, or less than, the dynamical timescale. This
is usually achieved in the simulations by setting the initial
conditions appropriately. In reality, this might be achieved either
via a
sudden change in the cooling rate (e.g. Johnson and Gammie 2003)
and/or by a sudden change in the disc density brought about by a
dynamical interaction with a low-mass interloper. Note that the mass
of such an interloper need only be comparable to the mass of the
proto-planetary disc (for example a massive planet or a brown dwarf),
and that such collisions might be frequent in the crowded and chaotic
conditions in which most stars form (Bate, Bonnell \& Bromm 2002a,
2002b, 2003).\footnote{Such ideas have resonance in the original ideas
of Jeans (1928), still being explored, for example, by Oxley \&
Woolfson (2004).} Black (1997) has noted that the similarity in
eccentricity distributions between exo-planets and close binary stars
which might indicate that both sets of formation processes are
dynamical in origin. There is indeed evidence that planetary
properties depend on stellar multiplicity (Zucker \& Mazeh 2002;
Eggenberger, Udry \& Mayor 2003). And such a dynamical mechanism for
forming planets is likely to give rise to a wide spread in
eccentricities and should be able to give rise to planets at a wide
range of orbital radii (Papaloizou \& Terquem 2001; Terquem \&
Papaloizou 2002).

In light of the above discussion, it is possible to take the view
that there are two distinct formation mechanisms for the formation of
planets. First, there is the well-trodden path for the formation of
the solar system, which produces rocky cores for gaseous planets at
radii of several AU as well as terrestrial planets at smaller radii,
perhaps on essentially circular orbits. This mechanism would also be
able to produce the debris discs at several tens to hundreds of
AU. Second, there is the possibility of a more dynamical mechanism for
planetary formation, using gravity directly in the formation
process. Such a mechanism might be able to produce gaseous planets at
a large range of radii, including the observed giant planets at radii
less than a few AU, and with a spread of orbital eccentricities. But
this mechanism would have difficulty in producing terrestrial-type
planets and debris discs. If this mechanism predominates among the
observed planetary systems, then few, if any, of the planetary systems
detected so far would harbour earth-like rocky planets.

These arguments are reinforced by the following point.  Most planetary
systems have non-negligible eccentricities, and there is a growing
body of opinion that habitability is not very sensitive to orbital
eccentricity (Williams \& Pollard 2002, 2003; Jones 2003;
Jones \& Sleep 2003; Asghari et al. 2004). So why, if the solar system
is not special, do its planets have essentially circular orbits? One
explanation for this might be that terrestrial planets must form
through steady core growth in a disc, and such a planet formation
mechanism tends to produce circular orbits.

A further piece of evidence that requires explanation is the
observation that there appears to be a correlation between the
metallicity of a star and the probability of it harbouring a planetary
system (Gonzalez, 1998; Gonzalez, Wallerstein \& Saar 1999; Reid
2002; Santos et al. 2003; Fischer \& Valenti 2003). Fischer \&
Valenti (2003, and private communication) find that the probability of
a star harbouring a planet rises from about 5 per cent when the iron
abundance is $\sim 1/3$ that of the sun, to about 20 per cent when the
iron abundance is $\sim3$ times that of the sun.

It could be argued (see, for example, Santos et al. 2003) that this
observed correlation can be taken as {\it prima facie} support for a
model of planet formation which requires the formation of rocky
planetary cores as a pre-requisite for forming the gas giants which we
observe. The correlation might be an indication that the formation
process is a continuous function of metallicity. In this picture, the
metal-rich systems would form planetary cores (and hence gas giants)
in abundance while there is still enough gas in the disc to cause
significant migration, somehow disposing of any excess cores so that
there is no debris left. The less metal-rich systems (including
the solar system) would be able to form a few cores which barely migrate
just as the gas is expelled, leaving a few gas giants at around $\sim$10
AU and a debris disc at larger radii.\footnote{We include the solar
  system in this category, although we note that its debris disc of
  around $10^{-5}$\,M$_{\earth}$ is somewhat smaller than for example
  the debris disc of $5\times 10^{-4}$\,M$_{\earth}$ found around the
  nearby sun-like (both in age and spectral type) star $\tau$ Ceti
  (Greaves et al. 2004b).}

In contrast to this rocky-core formation model, formation through
dynamical instability of a disc could lead to the observed giant
planets in small and eccentric 
orbits. Greaves et al. (2004a) find that these systems do not
show debris discs. In addition, they find that systems with observed
debris discs have suspected gas giants in large orbits (tens of
AU). Although systems like the solar system may
require some metallicity, this demonstrates the observed exo-planetary 
systems tend to be favoured by an even higher degree of
metallicity even though this is not a requirement of the disk
instability model. Indeed, Santos,
Israelian \& Mayor (2001) note that the `Sun occupies a modest position
in the low [Fe/H] tail of the metallicity distribution of stars with
planets' and go on to suggest that the lack of solar system analogues
found so far leads one `to speculate about possible different
formation histories'. 

From a theoretical point of view, there are many reasons, other than
core formation, as to why the formation of currently observable
planetary systems might be dependent on metallicity. For example,
cooling processes in the disc at low temperatures depend mainly on
metals (e.g. Johnson \& Gammie 2003), as does the shielding of the
disc from the X-ray flux generated in the stellar corona/chromosphere
(e.g. Alexander, Clarke \& Pringle 2004), as well as the cooling rate
in the region of the disc subject to evaporation (Hollenbach et al
1994). Thus, although metal abundance appears to play only a small
r\^ole in the disc timescale for viscous evolution (Livio \& Pringle
2004), it may play a significant r\^{o}le in the disc dynamics and/or
in the timescale on which the disc is dispersed by the central
star. In addition, the extra cooling brought about by enhanced metal
abundance might mean that the process of star formation takes place in
denser environments in more metal-rich systems, perhaps increasing
brown dwarf production, and in any case, enhancing the likelihood of
close interactions.

\section{Conclusions}

There are two main formation scenarios envisaged for planet
formation. The model applied to the solar system assumes that planets
form initially through the agglomeration of dust into rocks and
planetary cores, and that finally (for the giant planets) these cores
use gravity to accrete gas. Alternatively, there is the possibility of
a more dynamical mechanism using gravity directly in the formation
process. The solar system model may not be applicable to the observed
extrasolar planets and the dynamical instability model has difficulty
reproducing the solar system and debris discs. The metallicity
dependence of both models needs to be investigated further.

We conclude that it is still possible that our current understanding
of planetary systems is unduly coloured by our intimate knowledge of
our own solar system. More observational work is needed if the solar
system is to be shown to be a `normal' planetary system. And more
theoretical work is required if alternative planet formation scenarios
are to shown to be equally viable.

\section*{Acknowledgements}
JEP thanks STScI for hospitality and support from the Visitors'
Program. ML acknowledges support from John Templeton Foundation Grant
938-COS191. MEB acknowledges the support of a UKAFF fellowship and ARK
a Royal Society Wolfson Research Merit Award.

\bsp

\label{lastpage}
\end{document}